\documentclass[12pt]{article}

\begin{document}

\author{Sascha Vongehr \\
Department of Physics and Astronomy\\
University of Southern California\\
Los Angeles, CA 90089-0484\\
USA}
\title{Examples of Black Holes in Two-Time Physics }
\date{7.10.1999}
\maketitle

\begin{abstract}
Two time theory is derived via localization of the global Sp(2) symmetry [or
Osp(1/2), Osp(N/2), Sp(2N),...] in phase space in order to give a
self contained introduction to two time theory. Then it is shown that from
the two-times physics point of view theories of point particles on many
known black hole backgrounds are Sp(2) gauge duals of one another and of
course also gauge dual to all other equal dimensional gauges from earlier
two time related publications (hydrogen atom, ...). We reproduce the free
(quantum) relativistic particle on 1+1 dimensional black hole backgrounds
and 2+1 dimensional BTZ ones. Other 2+1 black holes and n+1 ones are
touched on but explicitely found only as cross sections of complicated
(n+1)+1 backgrounds. Further we give near horizon solutions (e.g. n+1
Robertson-Bertotti). Since two time physics can reproduce these backgrounds
all particle actions have hidden symmetries that oftentimes have not been noticed
before or whose origin was unclear. 
\end{abstract}

\newpage

\section{Two Time Theory}

\subsection{Introduction}

Early attempts of using two times go back as far as 1936 when Paul Dirac rederived $d=3+1$
Maxwell theory from a manifestly conformally invariant $d=4+2$ conformal space \cite{dirac}.
In order to remove the additional dimensions constraints had to be imposed artificially. 
We will derive two time theory from a completely different point of view. The need of two 
times follows rather being the starting point and the constraints are supplied by the theory itself. 

The huge success of today's fundamental theories like the standard model and
general relativity is due to localizations of global gauge freedoms. There
is one global symmetry that has never been localized. It is the $Sp(2)$
symmetry in phase space. Just like one arrives at general relativity via
localization of the Poincare symmetry, localizing the $Sp(2)$ symmetry in
phase space tells us that we need two times. Here we will derive two time
theory from this point of view with a formalism \cite{von} that facilitates
generalizations. We will show this for the introduction of spin. All that
will be needed is to substitute the $Sp(2)$ group metric with the $Osp(1/2)$
group metric. For the $Sp(2)$ case we will derive here for the first time
black hole backgrounds from two time theory. The derivation of two time
theory and the finding of examples - here black holes - should serve as a
self contained introduction into two time physics. We will restrict us to
bosonic examples, i.e. consider only $Sp(2)$ duals. For the motivations of
two time theory please see \cite{von} where we show as well motivations due
to considerations of M-theory. Only two time theory can give a large enough
supergroup to embed M-theory, thus two time theory should lead to M-theory
but there is only a toy-model \cite{bdmM} yet.

\subsection{Derivation of Two Time Theory}

\subsubsection{ The General Formalism}

For a group with metric $g$ and group elements $J$ acting on a multiplet $%
\Phi $ according to $\Phi ^{\prime }=J\Phi $ (or $\Phi _{j}^{\prime
}=J_{j}^{i}\Phi _{i}$) the following is an invariant:

\begin{equation}
\Phi ^{\dagger ^{\prime }}g\Phi ^{\prime }=\Phi ^{\dagger }g\Phi
\end{equation}

$\Phi ^{\dagger }\neq \Phi $ since one is a row and the other one a 
column but it will turn out that $\Phi _{i}^{\dagger }=\pm \Phi_{i}$.  
For infinitesimal transformations we can write $J=E+g\omega $ where $E$ is
the unit matrix and $\omega $ is called the generator. If $\Phi $ depends on
a parameter $\tau $ but $J$ may not then this is a global symmetry. In order
to make the symmetry local one introduces the covariant derivative $;_{\tau
}=(,_{\tau }-gA)$ where $A$ is the gauge field and $g$ still the group
metric (We do not consider a coupling strength.).

We choose the following for the group metric 
\begin{equation}
g^{\dagger }=g^{-1}=-g\Rightarrow g^{2}=-E  \label{groupmetriccondition}
\end{equation}
in order to ensure an anti-Hermitian norm and

\begin{equation}
\omega =\omega ^{\dagger }\Rightarrow A^{\dagger }=A
\end{equation}

We introduce the simplest (and well known) expressions that one could write down 
for an action and the Lagrangian in order to get the dimension of energy 
later on when $\Phi$ includes $X$ and $P$:

\begin{equation}
S=\int_{0}^{T}\pounds d\tau \quad \quad \quad \pounds =\frac{1}{2}\Phi
^{\dagger };_{\tau }g\Phi =\frac{1}{2}\Phi ^{\dagger },_{\tau }g\Phi -%
\frac{1}{2}\Phi ^{\dagger }A\Phi
\end{equation}

The Euler Lagrange equations give the equations of motion as: 
\begin{eqnarray}
\Phi ,_{\tau } &=&gA\Phi \Rightarrow \pounds ^{\dagger }=\pounds \\
\Phi _{i}\Phi _{j} &=&0  \label{constrainteom}
\end{eqnarray}

The latter tells us that we will only get trivial solutions if we do not
write 
\begin{equation}
\Phi _{i}\Phi _{j}=\Phi _{i}^{M}\Phi _{j}^{N}\eta _{MN}=0
\end{equation}
where $\eta $ has at least two times and is therefore the $SO(d,2)$ metric.
For clarity we will see all this with an example:

\subsubsection{$Sp(2)$ - Example}

$Sp(2)$ satisfies the condition (\ref{groupmetriccondition}). The symmetry
acts on the doublet 
\begin{equation}
\left( 
\begin{array}{c}
\Phi _{1} \\ 
\Phi _{2}
\end{array}
\right) =\left( 
\begin{array}{c}
X \\ 
P
\end{array}
\right) 
\end{equation}
where $(g_{ij})=\left( 
\begin{array}{cc}
0 & -1 \\ 
1 & 0
\end{array}
\right) $and $(g^{ij})=\left( 
\begin{array}{cc}
0 & 1 \\ 
-1 & 0
\end{array}
\right) $. The constraint (\ref{constrainteom}) translates into 
\begin{equation}
\Phi _{i}\Phi _{j}=0\Rightarrow X\cdot X=0=X\cdot P=0=P\cdot P
\end{equation}
Thus it follows that $X$ and $P$\ are two light like vectors that are not
parallel. Therefore they must have two time like dimensions and they must
have $d+2$ entries. We write $X^{M}$ where $M\in \{0,0^{\prime },1,...\}$
and $X^{M}$ is now a $SO(d,2)$ vector. Thus we see via the localization that
nature insists on two times because this is the only way not to get a
trivial theory after gauge fixing all three gauge choices that $Sp(2)\simeq
Sl(2,R)\simeq SO(2,1)$ offers. Every gauge choice means one choice of
functional
dependence of one of the $\Phi_{M}^{i}$ in terms of the others or put to equal 
a constant for example. Note that the two times do not come from a special 
choice of Lagrangian. We just used the simplest Lagrangian that gives us an energy
$X,_{\tau} P$. Taking the localisation of the $Sp(2)$ of ordinary physics 
seriously is leading to the two times. 

The three gauge choices fix a gauge surface inside the $SO(d,2)$ symmetric
starting space such that we are left with $SO(d-1,1)$. The third gauge
choice fixes the parametrization of the world line $\tau _{\left( t\right) }$%
. Note that we must have one more time and one more spatial direction. We
write $X^{M}$ where $M\in \{0,0^{\prime },1,1^{\prime },2,3,...,d-2\}$ . The
''accident'' $Sp(2)\simeq SO(2,1)$ means that we can interpret the above as
conformal gravity on the world line. We rewrite the quantization $\left[ X,P%
\right] =i\hbar $ as

\begin{equation}
\Phi _{i}g^{ij}\Phi _{j}=i\hbar  \label{quantization}
\end{equation}

\subsubsection{ The General Formalism Again}

$\eta _{MN}$ could be $diag(-1,-1,1,...,1)$ or related. The global $%
SO(d,2)$ contains d-dimensional Poincare symmetry $ISO(d-1,1)$ but there is
no translation invariance in $d+2$ dimensions. The $SO(d,2)$ generators are
due to Noether's theorem 
\begin{equation}
J^{MN}=\frac{1}{4}\Phi _{i}^{[M}g^{ij}\Phi _{j}^{N]}
\end{equation}

For the Casimirs holds classically 
\begin{equation}
C_{n[SO(d,2)]}=\frac{1}{n!}Tr\left( iJ\right) ^{n}\propto \Phi ^{M}\Phi
_{M}=0
\end{equation}
,that means that for all gauges one might choose later on the Casimir
invariants specify a unique representation of $SO(d,2)$. It completely
characterizes the gauge invariant physical space. The same holds for the
quantized theory but with $C_{n[SO(d,2)]}\neq 0$.

\subsubsection{$Osp(1/2)$ - and $Osp(N/2)$ - Example}

$Osp(1/2)$ satisfies the condition (\ref{groupmetriccondition}), too. The
symmetry acts on 
\begin{equation}
\left( 
\begin{array}{c}
\Phi _{1} \\ 
\Phi _{2} \\ 
\Phi _{3}
\end{array}
\right) =\left( 
\begin{array}{c}
X \\ 
P \\ 
\Psi 
\end{array}
\right) 
\end{equation}
where $(g_{ij})=\left( 
\begin{array}{ccc}
0 & -1 & 0 \\ 
1 & 0 & 0 \\ 
0 & 0 & i
\end{array}
\right) $and $(g^{ij})=\left( 
\begin{array}{ccc}
0 & 1 & 0 \\ 
-1 & 0 & 0 \\ 
0 & 0 & i
\end{array}
\right) $. The constraint (\ref{constrainteom}) translates into 
\begin{equation}
\Phi _{i}\Phi _{j}=0\Rightarrow X\cdot X=X\cdot P=P\cdot P=X\cdot \Psi
=P\cdot \Psi =0
\end{equation}
We have now five constraints matching the five parameters of $Osp(1/2)$. We
can interpret the above as conformal supergravity on the world line.
Introducing another arbitrary multiplet $\Theta _{i}$ the scalar product $%
\Phi g\Theta $ should be Hermitian and with $\Phi ^{\dagger }\neq \Phi$ 
(one is a row, the other one a column) and for the components
\begin{equation}
\Phi _{i}^{\dagger }=\pm \Phi _{i}  \label{plusminus}
\end{equation}
follows that $\Psi $ is Grassmanian. Thus, the quantization $\Phi
_{i}g^{ij}\Phi _{j}=i\hbar $ (\ref{quantization}) leads to $\left[ X,P\right]
=i\hbar $ again. The $\Phi _{i}^{\dagger }=\pm \Phi _{i}$ in (\ref{plusminus}%
) can be shown to be $\Phi _{i}^{\dagger }=+\Phi _{i}$ (It follows from the
properties of the generator $\omega $ ).

One might like to generalize to $Osp(N/2)$ using $\Psi _{a}^{M}$ .
Straightforward calculation results in: 
\begin{eqnarray}
J^{MN} &=&X^{M}P^{N}-P^{N}X^{M}+\frac{1}{2i}\sum_{a}\left( \Psi
_{a}^{M}\Psi _{a}^{N}-\Psi _{a}^{N}\Psi _{a}^{M}\right)  \\
\pounds  &=&\dot{X}\cdot P-\frac{i}{2}\sum_{a}\stackrel{\cdot }{\Psi }%
_{a}\Psi _{a}-\frac{1}{2}A^{ij}\Phi _{i}\Phi _{j}-\frac{1}{2}A^{ab}\Psi
_{a}\Psi _{b}+A^{ai}\Phi _{i}\Psi _{a}
\end{eqnarray}
where $i\in \{1,2\}$ . The generators show a spin and an angular momentum
part $J^{MN}=L^{MN}+S^{MN}$. The first order form for the Lagrangian has
been derived via dropping of total derivatives. It holds 
\begin{eqnarray}
\pounds ,_{\stackrel{\cdot }{X}} &=&P \\
\pounds ,_{\stackrel{\cdot }{\Psi }} &=&-\frac{i}{2}\Psi 
\end{eqnarray}
,the latter being a second class constraint.

\subsubsection{The General Formalism Again}

Starting with the general formalism one only needs to specify the group
metric $g$ in order to get different theories. We had $Sp(2)$ for a simple
point particle, $Osp(N/2)$ for a spinning particle but we could use for
example $Sp(2N)$ for a $N$ - particle system.

As we will see with the help of examples, the described theory is a
consistent and unitary quantum theory. The $SO(d,2)$ covariant quantization
gives the Casimir which for the bosonic point particle theory for example is 
$C_{2\left[ SO(d,2)\right] }=4C_{2\left[ Sp(2)\right] }-\frac{1}{4}(d^{2}-4)
$ where $C_{2\left[ Sp(2)\right] }=0$ because the physical states are
singlets in the gauge invariant sector. Thus we get 
\begin{equation}
C_{2\left[ SO(d,2)\right] }=-\frac{1}{4}(d^{2}-4)
\end{equation}
and the canonical or field theoretical quantization after gauge fixing
always gives $C_{2}=1-\left( \frac{d}{2}\right) ^{2}$ also \cite{ib}. This
generalizes according to the group used. For instance $Osp(N/2)$ leads to 
\begin{equation}
C_{2\left[ SO(d,2)\right] }=\frac{1}{8}(d+2)\left( N-2\right) \left(
d+N-2\right) 
\end{equation}

There have been discussions in the past \cite{m} \cite{s} \cite{mart} \cite{k}
of formalisms with two times, but not including the non-trivial classical and 
quantum solutions of two time physics \cite{ib}. \cite{m} suggested an action 
which can be gotten from our first order formalism by integrating out the momenta
$P$ in the path integral or semi-clasically by using its equation of motion. 
That second order formalism was obtained with different reasoning and motivation 
\cite{m} , and without the concept of gauge duality.

\section{$1+1$ Dimensional Black Hole Gauges}

The $1+1$ black hole is a solution of two-time theory in 2+2 dimensions by
choosing the $Sp(2,R)$ gauges $X^{+^{\prime }}=1$, $P^{+^{\prime }}=0$
(the third gauge is not yet fixed) and
solving the constraints $X^{2}=X\cdot P=0$ as follows 
\begin{eqnarray}
M &=&\left( +^{\prime }\quad -^{\prime }\quad +\quad -\right) \\
X^{M} &=&\left( 1,\quad -uv,\quad u,\,\quad v\right) N_{(u,v)}  \nonumber \\
P^{M} &=&\frac{1}{N}\left( 0,\quad up_{u}+vp_{v},\quad -p_{v},\quad
-p_{u}\right) \\
dX^{M} &=&\left( 0,\quad -d\left( uv\right) ,\quad du,\quad dv\right) N+%
\frac{X^{M}}{N}dN
\end{eqnarray}
The metric is given by $\eta ^{+^{\prime }-^{\prime }}=\eta ^{+-}=-1$ and
the line element is 
\begin{equation}
(ds)^{2}=(dX^{M})(dX_{M})=-2N^{2}dudv
\end{equation}
Inserting these forms in the original Sp$\left( 2,R\right) $ local and SO$%
\left( 2,2\right) \,$global invariant Lagrangian \cite{ib}\ gives 
\begin{eqnarray}
\pounds &=&\dot{X}\cdot P-\frac{1}{2}A^{22}P\cdot P-\frac{1}{2}A^{11}X\cdot
X-A^{12}X\cdot P \\
&=&\dot{u}p_{u}+\dot{v}p_{v}+\frac{A^{22}}{N^{2}}p_{u}p_{v} \\
&=&\frac{-N^{2}}{A^{22}}\dot{u}\,\dot{v}=\frac{1}{2A^{22}}G_{\mu \nu }\dot{x}%
^{\mu }\dot{x}^{\nu }
\end{eqnarray}
The metric is recognized in the line element or in the last line $G_{\mu \nu
}=\eta _{\mu \nu }N^{2}$, which is obtained by integrating out the momenta.
This Lagrangian describes a particle moving in the background of the black
hole given a suitable $N$.

This form shows that the system has the larger symmetry SO$\left( 2,2\right) 
$ whose generators are the Lorentz generators in the 2+2 dimensional space $%
L^{MN}=X^{M}P^{N}-X^{N}P^{M}$. In the present gauge these take a form that
is quantum ordered already: 
\begin{eqnarray}
L^{+^{\prime }-^{\prime }} &=&up_{u}+vp_{v},\quad L^{+^{\prime }+}=-p_{v} \\
L^{+^{\prime }-} &=&-p_{u},\quad L^{-^{\prime }+}=-u^{2}p_{u} \\
L^{-^{\prime }-} &=&-v^{2}p_{v},\quad L^{+-}=-up_{u}+vp_{v}
\end{eqnarray}
Under the SO$\left( 2,2\right) =SL\left( 2,R\right) _{L}\otimes SL\left(
2,R\right) _{R}$ the generators may be reclassified in the form 
\begin{eqnarray}
G_{2}^{L} &=&vp_{v},\quad G_{+}^{L}=G_{0}^{L}+G_{1}^{L}=-p_{v},\quad
G_{-}^{L}=G_{0}^{L}-G_{1}^{L}=-v^{2}p_{v}\,\,, \\
G_{2}^{R} &=&up_{u},\quad G_{+}^{R}=G_{0}^{R}+G_{1}^{R}=-p_{u},\quad
G_{-}^{R}=G_{0}^{R}-G_{1}^{R}=-u^{2}p_{u}\,\,.
\end{eqnarray}
where $G_{0}^{L,R}$ are the compact generators and $G_{1,2}^{L,R}$ are the
non-compact ones.

The classical SO$\left( 2,2\right) $ symmetry transformations generated by
these are obtained by evaluating the Poisson brackets $\delta u=\frac{1}{2}%
\varepsilon _{MN}\left\{ L^{MN},u\right\} $, $\delta v=\frac{1}{2}%
\varepsilon _{MN}\left\{ L^{MN},v\right\} $, which give 
\begin{eqnarray}
\delta u &=&-\varepsilon _{+^{\prime }-^{\prime }}u+\varepsilon _{+^{\prime
}+}0+\varepsilon _{+^{\prime }-}+\varepsilon _{-^{\prime
}+}u^{2}+\varepsilon _{-^{\prime }-}0+\varepsilon _{+-}u, \\
\delta v &=&-\varepsilon _{+^{\prime }-^{\prime }}v+\varepsilon _{+^{\prime
}+}+\varepsilon _{+^{\prime }-}0+\varepsilon _{-^{\prime }+}0+\varepsilon
_{-^{\prime }-}v^{2}-\varepsilon _{+-}v.
\end{eqnarray}
The Lagrangian density transforms as follows

\begin{equation}
\delta \left( \dot{u}\,\dot{v}N^{2}\right) =\dot{u}\,\dot{v}\Lambda \left(
\varepsilon ,u,v\right) 
\end{equation}
This transformation is cancelled by 
\begin{equation}
\delta \left( \frac{1}{A^{22}}\right) =\frac{-\Lambda \left( \varepsilon
,u,v\right) }{A^{22}}  \label{before}
\end{equation}
So that the Lagrangian is invariant under SO$\left( 2,2\right) $%
\begin{equation}
\delta \pounds =0.
\end{equation}
This larger symmetry of the particle action in for example the SL$\left(
2,R\right) /R$ black hole was not noticed before.

\subsection{Example: The SL$\left( 2,R\right) /R$ Black Hole Gauge}

The SL$\left( 2,R\right) /R$ black hole is described in terms of the Kruskal
coordinates by the line element 
\begin{equation}
ds^{2}=-\frac{du\,dv}{1-uv}
\end{equation}
The singularity is at $uv=1$. The region outside of the horizon is given by $%
u\geq 0\geq $ $v$ or by $v\geq 0\geq u$. The region inside the horizon is
given by $u,v\geq 0$ and $1\geq uv\geq 0$. The forbidden region where
particle geodesics cannot penetrate classically is $uv>1$. The black hole
that emerges from the gauged WZW model SL$\left( 2,R\right) /R$ is
reproduced with the choice $N=\frac{1}{\sqrt{1-uv}\sqrt{2}}$. For lambda
follows 
\begin{equation}
\Lambda \left( \varepsilon ,u,v\right) =\frac{-2\varepsilon _{+^{\prime
}-^{\prime }}+\left( \varepsilon _{+^{\prime }+}u+\varepsilon _{+^{\prime
}-}v\right) -\left( \varepsilon _{-^{\prime }-}v+\varepsilon _{-^{\prime
}+}u\right) \left( -2+uv\right) }{1-uv}.
\end{equation}

\subsection{More Examples:}

All $1+1$ black holes are conformally related. In order to find other $1+1$
black holes or two dimensional sections of higher dimensional ones we need
only find the right $N$. Here we would like to show this using Schwarzschild
coordinates. Defining $\sqrt{2}u=t+r_{\ast }$and $\sqrt{2}v=t-r_{\ast }$ the
gauge becomes 
\begin{eqnarray}
M &=&\left( +^{\prime }\quad \quad -^{\prime }\quad \quad 0\quad \quad
1\right) \\
X^{M} &=&\left( 1,\quad \frac{1}{2}\left( r_{\ast }^{2}-t^{2}\right) ,\quad
t,\,\quad r_{\ast }\right) N_{(t,r)} \label{ssscoor} \\
P^{M} &=&\frac{1}{N}\left( 0,\quad r_{\ast }p_{\ast }-tp_{t},\quad
p_{t},\quad p_{\ast }\right) \\
(ds)^{2} &=&N^{2}[-(dt)^{2}+(dr_{\ast })^{2}]
\end{eqnarray}

(or $t=\tau$ and $p_{t} =\left| p_{\ast }\right| $ for all three gauge
choices). The well known black hole metric 
\begin{equation}
ds^{2}=-N^{2}(dt)^{2}+N^{-2}(dr)^{2}
\end{equation}

requires $dr_{\ast }=drN^{-2}$. Thus we need to integrate to get the right $%
r_{\ast }$ from the $N$ that we want. The right canonical conjugate to $r$
is $p_{r}=p_{\ast }N^{-2}$. Then we can write the gauge in terms of $r$ and $%
p_{r}$ via writing the gauge with $r_{\ast }(r)$ and $p_{\ast
}(p_{r})=p_{r}N^{2}$(For example:$P^{-^{\prime }}=$ $[r_{\ast
}(r)p_{r}N^{2}-tp_{t}]\frac{1}{N}$).

Examples:

\begin{tabular}{|l|l|l|}
\hline
Black Hole & $N$ & $r_{\ast }$ \\ \hline
BTZ & $\sqrt{\left( \frac{r}{l}\right) ^{2}-M}$ & $\frac{l}{2\sqrt{M}}\ln 
\frac{r-l\sqrt{M}}{r+l\sqrt{M}}$ \\ \hline
Reissner-N. & $\sqrt{1-\frac{r_{+}}{r}}\sqrt{1-\frac{r_{-}}{r}}$ & $r+%
\frac{r_{+}^{2}\ln \left( r-r_{+}\right) +r_{-}^{2}\ln \left(
r-r_{-}\right) }{r_{+}-r_{-}}$ \\ \hline
\end{tabular}

\subsection{The Quantum Theory}

We quantize via $[u,p_{u}]=i=[v,p_{v}]$. It turns out that all operators
(like $u,p,r,L^{MN}$) are already in the ordering that makes them Hermitian
relative to the field theoretical dot product. The field theoretical Hermiticity
does not imply $L^{\dag} = L$ but requires these properties when the
$L$ act
on fields. The Hermiticity is then $\left\langle \phi | L \psi \right\rangle =
\left\langle L \phi | \psi \right\rangle $ where the dot product is:
$\left\langle \phi | L \psi \right\rangle = - \frac{i}{2} \int d^{d-1} x
[\phi^{\ast} (L\psi ),_{0}  - \phi,_{0}^{\ast} L\psi ]$.

With the gauges written as
above all $L^{MN}=X^{M}P^{N}-X^{N}P^{M}$ will close correctly. For example
the $G_{2}^{L,R}$, $G_{+}^{L,R}$ and $G_{-}^{L,R}$ have the standard $%
SL\left( 2,R\right) _{L,R}$ commutation rules among themselves 
\begin{equation}
\left[ G_{2}^{L,R},G_{\pm }^{L,R}\right] =\pm iG_{\pm }^{L,R},\quad \left[
G_{+}^{L,R},G_{-}^{L,R}\right] =-2iG_{2}^{L,R}.
\end{equation}

The conformal factor $N$ does not change the flat space generators in $d=2$
as shown in \cite{ibcd}. For the field theoretical operators that act on
wave functions write $p_{u}=-i\partial _{u}$ and so on. The Casimir is $\
C_{2}[SO(d,2)]=0$ as it should be in two dimensions since $C_{2}=$ 
$1-\frac{d^{2}}{4}$ \cite{ib}. Physical states are defined via $%
\left\langle \phi \left| 2p_{u}p_{v}\right| \phi \right\rangle =0$. This is
the $P^{M}P_{M}=0$ constraint due to the Lagrange multiplier $A^{22}$ , here
weakly enforced onto the states. This does not break the $SO(d,2)$ symmetry
since one can check that for all the generators $L^{MN}$

\begin{equation}
\left\langle \delta p_{u}p_{v}\right\rangle =\left\langle \phi \left| \left(
L^{MN}\right) ^{\dagger }(p_{u}p_{v})-(p_{u}p_{v})L^{MN}\right| \phi
\right\rangle  \label{check}
\end{equation}

\subsubsection{Canonical Gauge}

Leaving field theory one could have operators naively Hermitian via the
changes $u\rightarrow u-\frac{i}{2p_{u}}$, $v\rightarrow v-\frac{i}{2p_{v}}$
and

\begin{eqnarray}
L^{+^{\prime }-^{\prime }} &\rightarrow &L^{+^{\prime }-^{\prime }}-i \\
L^{-^{\prime }+} &\rightarrow &L^{-^{\prime }+}+iu-\frac{1}{4p_{u}}%
=-up_{u}u-\frac{1}{4p_{u}} \\
G_{2}^{L/R} &\rightarrow &G_{2}^{L/R}-\frac{i}{2} \\
L^{-^{\prime}-} &\rightarrow &L^{-^{\prime
}-}+iv-\frac{1}{4p_{v}}=vp_{v}v-\,\,\frac{1}{%
4p_{v}}
\end{eqnarray}

Then the generator's algebras again close correctly and the Casimir is as
desired $C_{2}=0$ also.

\section{BTZ Black Hole Gauges}

The BTZ black hole \cite{btz} is a solution of 2+1 Gravity with negative
cosmological constant $\Lambda =-\frac{1}{l^{2}}$ . Its line element can be
written 
\begin{equation}
ds^{2}=-N^{2}dt^{2}+N^{-2}d\rho ^{2}+\rho ^{2}\left( N^{\phi }\,dt+d\phi
\right) ^{2}  \label{A}
\end{equation}
where we work with $(8Gl)=1$ for convenience. Further holds 
\begin{eqnarray}
l^{2}N^{2} &=&\left( \rho ^{2}-Ml^{2}+\left( \frac{Jl}{2\rho }\right)
^{2}\right) \\
N^{\phi } &=&-\frac{J}{2\rho ^{2}}
\end{eqnarray}
and the outer/inner horizon is at 
\begin{eqnarray}
\rho _{\pm }^{2} &=&\frac{1}{2}\left( Ml^{2}\pm r_{+}^{2}\right)  \label{B}
\\
r_{+}^{2} &=&Ml^{2}\sqrt{1-\left( \frac{J}{Ml}\right) ^{2}}
\end{eqnarray}

The AdS gauge of \cite{ibcd} can not only be used to describe the global AdS
case but also for backgrounds that are locally AdS only. The covering space
of AdS space is a BTZ solution with negative mass and the standard method 
\cite{djg} needs analytic continuations, that is it allows imaginary numbers 
as values for angles and times for instance,  in order to yield
BTZ black holes from the AdS covering space. This can now be understood in a
more natural way as a $Sp(2)$ duality transformation in two time physics 
\cite{ib}.

We start out with (124) to (126) of \cite{ibcd}:
\begin{eqnarray}
M     &=&\left( \,0^{\prime }\quad ,\quad \,1^{\prime }\quad ,\quad \,m
\right) \\
X^{M} &=&\left( \pm \sqrt{1+X_{m}^{2} (x)} \,\quad ,1 \, \quad ,X^{m} (x)
\right) \label{import} \\
P^{M} &=&\left( \frac{X^{m} (x) e_{m}^{\mu } (x) p_{\mu }}{\pm
\sqrt{1+X_{m}^{2} (x) } } \, \quad ,0 \, \quad ,e_{m}^{\mu } (x) p_{\mu }
\right) 
\end{eqnarray}
where $e_{m}^{\mu}$ is the inverse  of $e_{\mu}^{m}$ and designed just 
such that $p_{\mu}$ has the meaning of canonical momentum:
\begin{eqnarray}
e_{\mu }^{m} = X,_{\mu}^{m} (x) - X^{m} (x) \frac{X_{n} (x) X,_{\mu}^{n}}{1+X^{2}_{m} (x)} \label{import2}
\end{eqnarray}

Note that $P^{2}=0$ has not been imposed yet, and there still is one 
more gauge freedom. This we rewrite: 
\begin{eqnarray}
M &=&\left( \,0^{\prime }\quad \,1^{\prime }\quad \,0\quad \,1\quad
\,2\,\right) \\
X^{M} &=&\left( C\gamma ,\quad C\sigma ,\quad Ss,\,\quad Sc,\quad 1\right) l
\end{eqnarray}
where $C^{2}-S^{2}=\mp 1$ , $c^{2}-s^{2}=\pm 1$ and $\gamma ^{2}-\sigma
^{2}=1$ such that $X^{M}X_{M}=0$. The upper sign will be the solution
outside the event horizon and the lower sign will give the inside solution. $%
dX^{M}dX_{M}$ is the line element 
\[
ds^{2}=\left( \mp S^{2}\left( \frac{ds}{c}\right) ^{2}\pm \left( \frac{dC}{S}%
\right) ^{2}+C^{2}\left( \frac{d\sigma }{\gamma }\right) ^{2}\right) l^{2} 
\]
In order to see the BTZ black hole in this reparametrize as follows: 
\begin{equation}
c=\cosh \frac{x^{+}}{l}\Longleftrightarrow s=\sinh \frac{x^{+}}{l}
\end{equation}
for the outside solution and 
\begin{equation}
c=\sinh \frac{x^{+}}{l}\Longleftrightarrow s=\cosh \frac{x^{+}}{l}
\end{equation}
for the solution inside the event horizon. Further we pick 
\begin{equation}
\gamma =\cosh \frac{x^{-}}{l}\Longleftrightarrow \sigma =\sinh \frac{x^{-}}{l%
}
\end{equation}
It follows 
\begin{equation}
ds^{2}=\mp S^{2}\left( dx^{+}\right) ^{2}\pm S^{-2}l^{2}(dC)^{2}+C^{2}\left(
dx^{-}\right) ^{2}  \label{le}
\end{equation}

Let us concentrate from now on onwards on the outside solution just to avoid
the proliferation of $\pm $. Then $C,c,\gamma $ are or at least behave like $%
\cosh $ functions in many respects and $S,s,\sigma $ like $\sinh $ ones. Let 
$C,S$ be only dependent on a parameter $R$. Using equations (\ref{import}) 
and (\ref{import2}) yields the canonical momenta $p_{R},p_{\frac{x^{\pm }}{l}}$
which we will write in the combinations 
\begin{eqnarray}
P &=&\frac{C}{S,_{R}}p_{R}=\frac{S}{C,_{R}}p_{R} \\
p &=&\frac{C}{S}p_{\frac{x^{+}}{l}} \\
\pi  &=&\frac{S}{C}p_{\frac{x^{-}}{l}}
\end{eqnarray}
Therefore: 
\[
P^{M}=\frac{1}{l}\left( \left( S\gamma P+\frac{\sigma }{S}\pi \right)
,\left( S\sigma P+\frac{\gamma }{S}\pi \right) ,\left( CsP-\frac{c}{C}%
p\right) ,\left( CcP-\frac{s}{C}p\right) ,0\right) 
\]
In lightcone coordinates $X^{\pm }=X^{0}\pm X^{1}$ , i.e. $\eta ^{+^{\prime
}-^{\prime }}=-\frac{1}{2}=\eta ^{+-}$ we have: 
\begin{eqnarray}
M &=&\left( \pm ^{\prime }\quad ,\quad \pm \quad ,\quad 2\right)  \\
X^{M} &=&l\left( Ce^{\pm \left( \frac{x^{-}}{l}\right) },\pm Se^{\pm \left( 
\frac{x^{+}}{l}\right) },1\right)  \\
P^{M} &=&\frac{1}{l}\left( e^{\pm \frac{x^{-}}{l}}\left( SP\pm \frac{\pi }{S}%
\right) ,\pm e^{\pm \frac{x^{+}}{l}}\left( CP\mp \frac{p}{C}\right)
,0\right) 
\end{eqnarray}
Inserting these into the original SO$(3,2)$ action gives 
\begin{eqnarray}
L &=&\dot{X}^{M}P_{M}-\frac{A_{22}}{2}P^{M}P_{M}  \nonumber \\
&=&\frac{\dot{x}^{+}}{l}p_{\frac{x^{+}}{l}}+\dot{R}p_{R}+\frac{\dot{x}^{-}}{l%
}p_{\frac{x^{-}}{l}}  \nonumber \\
&&-\frac{A_{22}}{2l^{2}}\left( -S^{-2}p_{\frac{x^{+}}{l}}^{2}+\left( \frac{S%
}{C,_{R}}\right) ^{2}p_{R}^{2}+C^{-2}p_{\frac{x^{-}}{l}}^{2}\right) 
\end{eqnarray}
From the last bracket one can read off the metric 
\begin{eqnarray}
G^{\mu \nu } &=&\frac{1}{l^{2}}diag\left( -S^{-2},\left( \frac{S^{2}}{%
C^{2},_{R}}\right) ,C^{-2}\right)  \\
\Rightarrow G_{\mu \nu } &=&l^{2}diag\left( -S^{2},\left( \frac{C^{2},_{R}}{%
S^{2}}\right) ,C^{2}\right) 
\end{eqnarray}
which is consistent with the line element (\ref{le}).

\subsection{Classical Generators of the larger Symmetry}

The classical generators of the two time Lorentz symmetry according to $%
L^{MN}=X^{M}P^{N}-X^{N}P^{M}$ are as follows: 
\begin{eqnarray}
L^{2M} &=&lP^{M} \\
L^{0^{\prime }1^{\prime }} &=&p_{\left( \frac{x^{-}}{l}\right)
}=L^{+^{\prime }-^{\prime }} \\
L^{01} &=&p_{\left( \frac{x^{+}}{l}\right) }=L^{+-} \\
L^{0^{\prime }0} &=&\gamma sP-\gamma cp-s\sigma \pi  \\
L^{1^{\prime }1} &=&\sigma cP-\sigma sp-c\gamma \pi  \\
L^{0^{\prime }1} &=&\gamma cP-\gamma sp-c\sigma \pi  \\
L^{1^{\prime }0} &=&\sigma sP-\sigma cp-s\gamma \pi  \\
L^{\pm ^{\prime }+} &=&\frac{1}{2}exp\left( +\left( \frac{x^{+}}{l}\pm \frac{%
x^{-}}{l}\right) \right) \left( p\pm \pi -P\right)  \\
L^{\mp ^{\prime }-} &=&\frac{1}{2}exp\left( -\left( \frac{x^{+}}{l}\pm \frac{%
x^{-}}{l}\right) \right) \left( p\pm \pi +P\right) 
\end{eqnarray}
To split the SO$(2,2)$ subalgebras into $SL(2,R)_{L}\bigotimes SL(2,R)_{R}$
seems not helpful. Given the complexity of the operators, finding the
quantum anomalies with $\left[ R,p_{R}\right] =i$ etc. would be a tour de
force. The $SO(3,2)$ transformations generated by the $L^{MN}$ are again the
Poisson brackets $\delta =\frac{1}{2}\epsilon _{MN}\left\{ L^{MN},\,\right\} 
$ 
\begin{eqnarray}
\delta R &=&(-\epsilon _{0^{\prime }0}s\gamma -\epsilon _{0^{\prime
}1}c\gamma +\epsilon _{0^{\prime }2}S\gamma -\epsilon _{1^{\prime }0}s\sigma 
\nonumber \\
&&-\epsilon _{1^{\prime }1}c\sigma +\epsilon _{1^{\prime }2}S\sigma
+\epsilon _{02}Cs+\epsilon _{12}Cc)\frac{C}{S,_{R}} \\
\delta \left( \frac{x^{+}}{l}\right)  &=&(\epsilon _{0^{\prime }0}c\gamma
+\epsilon _{0^{\prime }1}s\gamma +\epsilon _{1^{\prime }0}c\sigma +\epsilon
_{1^{\prime }1}s\sigma )\frac{C}{S}  \nonumber \\
&&-\epsilon _{01}-\epsilon _{02}\frac{c}{S}-\epsilon _{12}\frac{s}{S} \\
\delta \left( \frac{x^{-}}{l}\right)  &=&(\epsilon _{0^{\prime }0}s\sigma
+\epsilon _{0^{\prime }1}c\sigma +\epsilon _{1^{\prime }0}s\gamma +\epsilon
_{1^{\prime }1}c\gamma )\frac{S}{C}  \nonumber \\
&&-\epsilon _{0^{\prime }1^{\prime }}+\epsilon _{1^{\prime }2}\frac{\gamma }{%
C}+\epsilon _{0^{\prime }2}\frac{\sigma }{C}
\end{eqnarray}
With $\dot{X}^{\mu }$ dependent on $R,x^{+},x^{-}$ the Lagrangian transforms
as 
\begin{equation}
\delta L=\delta (\frac{1}{2A_{22}}G_{\mu \nu }\dot{X}^{\mu }\dot{X}^{\nu })
\end{equation}
which gives together with the demand that $\delta L$ should be at most a
total derivative how $\delta A_{22}$ has to look like such that the action
is invariant - just like we did before (\ref{before}). This bigger SO$(3,2)$
symmetry of the particle action on a BTZ background was to our knowledge not
known.

\subsection{Example I: A Rotating Outside Solution with Arbitrary Mass}

Lets actually give examples for the C and S functions. For $C=\cosh R$
follows $S=\sinh R$ and therefore $P=p_{R}$. Hence the line element to be
compared with (\ref{le}) is 
\begin{equation}
ds^{2}=-S^{2}\left( dx^{+}\right) ^{2}+l^{2}(dR)^{2}+C^{2}\left(
dx^{-}\right) ^{2}
\end{equation}
which holds for the outside of the black hole and is connected to more
intuitive coordinates via $x^{\pm }=(\rho _{\pm }\frac{t}{l}-\rho _{\mp
}\phi )$. The equations (\ref{A}) to (\ref{B}) can now easily be reproduced;
i.e. this is a BTZ black hole with $\rho _{\pm }$ as inner and outer event
horizons and the mass and angular momentum given by the arithmetic and
harmonic means of $( \frac{\sqrt{2}}{l} \rho _{\pm } )^{2}$: 
\begin{eqnarray}
M &=&\frac{1}{l^{2}}(\rho _{+}^{2}+\rho _{-}^{2}) \\
J &=&\frac{2}{l}(\rho _{+} \rho _{-})
\end{eqnarray}

\subsection{Example II: Outside and Inside Solutions for the J=0 Black Hole
with Fixed Mass}

From the choice $C=\frac{\rho }{l}$ follows $S=\sqrt{\pm (\left( \frac{\rho 
}{l}\right) ^{2}-1)}$ ($+$ for $\rho \geq l$ , $-$ for $l\geq \rho $) and
thus $P=\sqrt{\rho ^{2}-l^{2}}$. To make contact with well known notation
define $t=x^{+}$ , $\phi =\frac{x^{-}}{l}$ and $\rho =R$ , such that
(compare (\ref{A})) 
\begin{eqnarray}
ds^{2} &=&-N^{2}dt^{2}+N^{-2}d\rho ^{2}+\rho ^{2}d\phi ^{2} \\
N &=&\sqrt{\left( \frac{\rho }{l}\right) ^{2}-1}
\end{eqnarray}
It is an infinitely leafed space time: $+\infty >\rho \geq 0$ and $-\infty
<t,\phi <+\infty $

An identification to $\phi \equiv \phi +2\pi $ is an identification of
points as well in the larger space time having two times.

\section{Other 2+1 and n+1 Backgrounds}

\subsection{Ansatz with Conformal Gauge}

The BTZ gauges feature rather involved operators that allow no straight
forward generalization to higher dimensions or, since they came from an
AdS-gauge, not to other curvatures. The conformal gauge of the $1+1$
solutions given earlier can easily be generalized (compare (\ref{ssscoor}) 
:

\begin{eqnarray}
M &=&\left( +^{\prime }\quad \quad -^{\prime }\quad \ \ \ \ \quad 0\quad \ \
\ \ \ \ \ \quad i\right) \\
X^{M} &=&\left( 1,\quad \frac{1}{2}\left( r_{\ast }^{2}-B^{2}\right) ,\quad
B,\,\quad r_{\ast }\widehat{\Omega }^{i}\right) \left( \frac{r}{r_{\ast }}%
\right)
\end{eqnarray}

where $\widehat{\Omega } $ is a unit vector.
Demanding the metric $(ds)^{2}=-N^{2}(dt)^{2}+N^{-2}(dr)^{2}+r^{2}(d\widehat{%
\Omega })^{2}$ needs $\stackrel{.}{B}B^{\prime }=\stackrel{.}{r_{\ast }}%
r_{\ast }^{\prime }$ in order to cancel the $(dt)(dr)$ term and the
conditions

\begin{equation}
(\stackrel{.}{B}^{2}-\stackrel{.}{r_{\ast }}^{2})\left( \frac{r}{r_{\ast }}%
\right) =N^{2}=\frac{r_{\ast }}{r\sqrt{r_{\ast }^{\prime 2}-B^{\prime 2}}}
\end{equation}

($\stackrel{.}{N}=0$ is understood). This leads to 
\begin{eqnarray}
\partial _{r}\left( N^{2}\partial _{r}B\right) -\partial _{t}\left(
N^{-2}\partial _{t}B\right) &=&0  \label{scalar} \\
\partial _{r}\left( N^{2}\partial _{r}r_{\ast }\right) -\partial _{t}\left(
N^{-2}\partial _{t}r_{\ast }\right) &=&0
\end{eqnarray}

, i.e. the solutions are scalar fields on the very backgrounds given via $N$
[recall $\nabla ^{2}\Phi =\partial _{\mu }\left( \sqrt{-G}G^{\mu \nu
}\partial _{\nu }\Phi \right) $]. Even for simple $N$ it leads to non linear
PDEs and all one could possibly gain are background metrics that can be put
into the conformal gauge form. Two time theory becomes especially rich with $%
X^{M}=X_{\left( x^{\mu },p^{\mu }\right) }^{M}$ but we restrict us to the
conformal-gauge ansatz because with $X^{M}=\left( ...\right) _{\left( x^{\mu
}\right) }N_{\left( x^{\mu }\right) }$ the metric can be found very easily.

\subsection{Near Horizon Gauges}

Sleight deviations from the conformal gauge ansatz above led to the
following near horizon backgrounds:

\subsubsection{Robertson-Bertotti Background in any Dimensions}

\begin{eqnarray}
M &=&\left( +^{\prime }\quad \quad -^{\prime }\quad \ \ \quad 0\quad \ \
\quad i\right)  \nonumber \\
X^{M} &=&\left( 1,\quad \frac{1}{2}\left( r_{\ast }^{2}-t^{2}\right) ,\quad
t,\,\quad r_{\ast }\widehat{\Omega }^{i}\right) N \\
P^{M} &=&\frac{1}{N}\left( 0,\quad r_{\ast }p_{\ast }-tp_{t},\quad
p_{t},\quad p_{\ast }^{i}\right) \\
p_{\ast }^{i} &=&p_{\ast }\left( \widehat{\Omega }^{i}+\alpha L^{ij}\widehat{%
\Omega }_{j}\right)
\end{eqnarray}

With $r_{\ast }=-\left( \frac{M^{2}}{r}\right) $ \ and $N=\left( \frac{r}{M}%
\right) $ \ follows 

\begin{equation}
(ds)^{2}=[-N^{2}(dt)^{2}+N^{-2}(dr)^{2}]+M^{2}(d%
\widehat{\Omega })^{2}
\end{equation}

\subsubsection{A d=2+1 Near Horizon Gauge}

\begin{eqnarray}
M &=&\left( \quad +^{\prime }\quad \quad \ \quad -^{\prime }\quad \ \ \quad
\quad \ 0\quad \ \ \quad 1\quad \ \ 2\right)  \nonumber \\
X^{M} &=&\left( 1,\quad \frac{\left( \frac{M^{2}}{r}\right)
^{2}-t^{2}+(M\phi )^{2}}{2},\quad t,\,\quad -\left( \frac{M^{2}}{r}\right)
,\quad M\phi \right) N \\
P^{M} &=&\frac{1}{N}\left( \quad 0,\quad -\phi p_{\phi }-\left( \frac{M^{2}%
}{r}\right) p_{r}N^{2}-tp_{t},\quad p_{t},\quad p_{r}N^{2},\quad \frac{%
p_{\phi }}{M}\right)
\end{eqnarray}

With $M$ a constant not dependent of $r$ and $N=\left( \frac{r}{M}\right) $
\ follows $(ds)^{2}=[-N^{2}(dt)^{2}+N^{-2}(dr)^{2}]+r^{2}(d\phi )^{2}$. The
Lagrangian $L=\dot{X}^{M}P_{M}-\frac{1}{2}A^{22}P^{M}P_{M}$ is 
\begin{equation}
L=-\stackrel{.}{t}p_{t}-\stackrel{.}{r}p_{r}-\stackrel{.}{\phi }p_{\phi }-%
\frac{A^{22}}{2}\left( -\left( \frac{p_{t}}{N}\right) ^{2}+\left(
Np_{r}\right) ^{2}+\left( \frac{p_{\phi }}{NM}\right) ^{2}\right)
\end{equation}

Again we would like to stress that the derivation of these systems - here
particles in a curved space time\ -\ as two time theory gauges proves they
have symmetries hidden in the action that have gone unnoticed or whose origin 
was clouded before and that make them $SO(d,2)$ symmetric systems.

\subsection{Backgrounds containing other Black Holes}

The following is to demonstrate that basically all black hole backgrounds
can be modelled with two time theory gauges. The following is the 
$\widehat{\Omega }^{1}=0 $ subspace of a more complicated gauge. 
\begin{eqnarray}
M &=&\left( \quad +^{\prime }\quad \quad \ \quad -^{\prime }\quad \ \ \quad
\quad \ 0\quad \ \ \quad 1\quad \ \ i>1\right)   \nonumber \\
X^{M} &=&\quad \left( 1,\quad \frac{r_{\ast }^{2}-t^{2}+\left( \frac{r}{N}%
\right) ^{2}}{2},\quad t,\,\quad r_{\ast },\quad \frac{r}{N}\widehat{\Omega }%
^{i}\right) N \\
P^{M} &=&\frac{1}{N}\left( \quad 0,\quad p_{r}r-tp_{t}-p_{\ast }r_{\ast
},\quad p_{t},\quad p_{\ast },\quad p_{r}N\left( \varpi \right) \right) 
\end{eqnarray}

Where $\left( \varpi \right) =\left( \widehat{\Omega }^{i}+\alpha L^{ij}%
\widehat{\Omega }_{j}\right) $. It follows 
\begin{equation}
(ds)^{2}=-N^{2}(dt)^{2}+N^{+2}(dr_{\ast })^{2}+N^{+2}(d\frac{r}{N}%
)^{2}+r^{2}(d\widehat{\Omega })^{2}
\end{equation}
such that black hole backgrounds - here non rotating ones - can be gotten
via integrating with the desired $N$ : 
\begin{equation}
r_{\ast }=\int dr\frac{1}{N^{2}}\sqrt{1-\left( N-rN^{\prime }\right) ^{2}}
\end{equation}

$N=\left( \frac{r}{M}\right) $ for example gives again $r_{\ast }=-\left( 
\frac{M^{2}}{r}\right) $. Examples for black holes are:

BTZ: 
\begin{eqnarray}
N &=&\sqrt{\pm \left[ \left( \frac{r}{l}\right) ^{2}-M\right] }\quad
\Longrightarrow \quad \left( N-rN^{\prime }\right) =\frac{M}{N} \\
&\Longrightarrow &r_{\ast }=\int dr\left[ \frac{1}{N^{2}}\sqrt{\pm \left[
1-\left( \frac{M}{N}\right) ^{2}\right] }\right]
\end{eqnarray}

$\pm $ are for the inside/outside solutions where for the latter one has to
exchange $X^{0}$ and $X^{1}$ and adjust $X^{-}$ and $P^{M}$ in order to
still satisfy the $X^{M}X_{M}=0=P^{M}P_{M}$ constraints.

Reissner-Nordstrom: 
\begin{eqnarray}
N &=&\left( N_{+}N_{-}\right) \quad \quad ;\quad N_{\pm }=\sqrt{1-\frac{%
r_{\pm }}{r}} \\
&\Longrightarrow &r_{\ast }=\pm \int dr\left[ \frac{1}{N}\sqrt{\frac{1}{N^{2}%
}-\left( 2-\frac{1}{2N_{+}^{2}}-\frac{1}{2N_{-}^{2}}\right) ^{2}}\right]
\end{eqnarray}

This is the outside solution. The center and inside solutions are again very
similar.

\section{Conclusions}

Many black hole backgrounds have been shown to be two time theory gauges.
Thereby we established $Sp(2)$ gauge duality between several backgrounds. 
We demonstrated that two time theory is certainly rich enough to allow for
and unify very many complicated systems. One should bear in mind that all of
the above was found with the help of either the AdS gauge or another rather
restricted ansatz (namely the conformal gauge) but two time theory is much
richer. As well, the $d+2$ gauges are dual to all the other gauges of equal
dimensionality that have been published before, as there are the hydrogen
atom and the harmonic oscillator \cite{ibhydro} and so on.

Particles in black hole backgrounds having a conformal metric and ones near 
event horizons show conformal symmetries. For $1+1$ and $2+1$ BTZ backgrounds
it was explicitly shown how this symmetry which is hidden in the action 
emerges from the $SO(d,2)$ that acts linearly in the ungauged two time space but is non
linearly realized once a gauge surface is fixed. As observed in \cite{ibt}
and \cite{von} already: The $AdS_{n}\times S^{m}$ discussion may benefit
from this - symmetries hidden in the action can be as vital as the Lorentz
boost symmetry for a free particle (A Lagrangian like $L=-m\sqrt{1-\dot{r}%
^{2}}$ or the Hamiltonian $H=\sqrt{p^{2}+m^{2}}$do not reveal it.).

The method of \cite{djg} has been given a natural justification. In order to
go from global AdS to the black hole solution one has had to identify 
(make time periodic) and complexify (use imaginary numbers for real 
quantities) because both are gauges in a space time with two times. The
difference between the gauges is how the gauge slices lie in the bigger
space time. The effective exchange of the roles of one of the times with
that of one space like direction relative to the gauge surface was formally
achieved via the complexification.

The methods used here often remind of purely geometrical embeddings in order
to obtain a metric more conveniently. Hence, we would like to stress that
all our gauge choices are consistent, dynamical quantum theories that allow
for introduction of spin \cite{ibcd}\ and supersymmetry \cite{bdm}.

\section{Acknowledgments}

I greatly benefited from many a discussion with Itzhak Bars. He also came up
with the SL$\left( 2,R\right) /R$ black hole gauge and made me understand
that one needs to check (\ref{check}) and use not only naive Hermiticity but
Hermiticity due to the field theoretical dot product. Itzhak was the first
to realize that the conformal gauge ansatz lead to (\ref{scalar}) whose
solutions are scalar fields on the very backgrounds given.


\begin{thebibliography}{9}
\bibitem{dirac}  P. Dirac, Ann.Math. 37: 429, (1936)

\bibitem{von}  S. Vongehr, ''Nature:''I have Two Times'' hep-th/9907034

\bibitem{bdmM}  I. Bars, C. Deliduman, D. Minic,''Lifting M-Theory to
Two-Time Physics'' hep-th/9904063

\bibitem{ib}  I. Bars, ''Two-Time Physics'' hep-th/9809034

\bibitem{m}  R. Marnelius, ''Manifestly Conformal Covariant Description 
of Spinning and Charged Particles'', Phys.Rev. D20 (1979) 2091

\bibitem{s}  W. Siegel, ''Conformal Invariance of Extended Spinning Particle
Mechanics'', Int.J.Mod.Phys. A3 (1988) 2713-2718

\bibitem{mart}  U. Martensson, Int.J.Mod.Phys. A8  (1993) 5305

\bibitem{k}  S.M. Kuzenko and J.V. Yarevskaya, Mod.Phys.Lett. A11 (1996) 1653

\bibitem{ibcd}  I. Bars and C. Deliduman,''Gauge symmetry in phase space
with spin'' hep-th/9806085, Phys.Rev. D58 (1998) 106004

\bibitem{btz}  M. Banados, C. Teitelboim and J. Zanelly, Phys.Rev.Lett. 69
(1992) 1849-1851

\bibitem{djg}  S. Deser, R. Jackiw and G't Hooft, Ann.Phys. 152 (1984) 220

\bibitem{ibt}  I. Bars, ''Hidden Symmetries, AdS\_D x S\symbol{94}n, and the
lifting of one-time-physics to two-time-physics'' hep-th/9810025, Phys.Rev.
D59 (1999) 045019

\bibitem{ibhydro}  I. Bars, ''Conformal Symmetry and Duality between Free
Particle, H-atom and Harmonic Oscillator'' hep-th/9804028, Phys.Rev. D58
(1998) 066006

\bibitem{bdm}  I. Bars, C. Deliduman, D. Minic, ''Supersymmetric Two-Time
Physics'' hep-th/9812161, Phys.Rev. D59 (1999) 125004
\end{thebibliography}
\end{document}